\documentclass[twocolumn]{elsart}
\usepackage{graphicx}
\usepackage{natbib}

\def\Rate{{\it R}}

\begin{document}
\runauthor{Kim, Kalogera, Lorimer, O'Shaughnessy, Belczynski}
\begin{frontmatter}
\title{Effect of PSR J0737-3039 on the DNS Merger Rate and Implications for GW Detection}
\author{Chunglee Kim,~}\ead{c-kim1@northwestern.edu}
\author{Vicky Kalogera,}

\address{Northwestern University, Department of Physics and Astronomy, 2145 Sheridan Rd., Evanston, IL 60208, USA}

\author{Duncan Lorimer}
\address{University of Manchester, Jodrell Bank Observatory, Macclesfield, Cheshire, SK11 9DL, UK}

\begin{abstract}
We present the current estimates of the Galactic merger rate of double-neutron-star (DNS) systems. Using a statistical analysis method, we calculate the probability distribution function (PDF) of the rate estimates, which allows us to assign confidence intervals to the rate estimates. We calculate the Galactic DNS merger rate based on the three known systems B1913+16, B1534+12, and J0737-3039. The discovery of J0737-3039 increases the estimated DNS merger rate by a factor $\sim$6 than what is previously known. The most likely values of DNS merger rate lie in the range $3-190$ Myr$^{-1}$ depending on different pulsar models. Motivated by a strong correlation between the peak rate estimates and a pulsar luminosity function, we calculate a {\em global} probability distribution as a single representation of the parameter space covered by different pulsar population models. We compare the global PDF with the observed supernova Ib/c rate, which sets an upper limit on the DNS merger rate. Finally, we remark on implications of new discoveries such as of J1756$-$2251, the 4th DNS in the Galactic disk, and J1906+0746, a possible DNS system.

\end{abstract}
\begin{keyword}
method : statistical ; stars : binary ; pulsars : individual (PSR J0737-3039) ; gravitational waves
\end{keyword}
\end{frontmatter}

\section{Introduction}
Merging (or inspiralling) DNS systems are one of the prime targets for ground-based gravitational-wave (GW) interferometers such as GEO600, TAMA, VIRGO, and LIGO (the Laser Interferometer Gravitational-Wave Observatory) \citep{ligo}. Event rates of the DNS inspiral searches by these detectors can be inferred using the rate estimates with an extrapolation out to the maximum detection distances for any detector under consideration. Before 2003, the Galactic DNS merger rate had been estimated between $\sim10^{-7} - 10^{-5}$ yr$^{-1}$ (see Kalogera et al.\ 2001\nocite{knst} and references therein). At that time, there were only two systems available for empirical studies, PSRs~B1913+16 \citep{ht75} and B1534+12 \citep{w91}. We calculated the PDF of Galactic DNS merger rates, $P(R)$, based on these two systems (see Kim et al.\ 2003, hereafter KKL, for further discussions). Soon after the discovery of the highly relativistic system PSR~J0737$-$3039 \nocite{0737A}, we were able to revise $P(R)$ including PSR~J0737$-$3039 in collaboration with the observation team\footnote{Only the millisecond component (J0737-3039A) is considered in our calculation. For instance, the current age of the system is derived from the pulsar A.}\citep{0737A, k04}. The discovery of J0737$-$3039, resulted in a significant  increase in the estimated Galactic DNS by a factor of $\sim$6. This implies a boost in event rates for DNS searches for GW interferometers. In this paper, we present our recent results on: (i) the PDF of Galactic DNS merger rate estimates with updated observations; (ii) the {\em global\/} PDF of rate estimates considering the systematic uncertainties; (iii) constraints on upper limits for rate estimates based on the observed supernova rate incorporated with our theoretical understanding of the SN-DNS relation; (iv) the approximate contribution of J1756$-$2251 to the Galactic DNS merger rates and uncertainties in our assumptions on the efficiency of the acceleration searches for Parkes multibeam pulsar survey (PMPS). Finally, we discuss implications of the most recently discovered pulsar binary J1906+0746 \citep{j1906}.

\section{The Galactic DNS merger rate}
Here, we describe the main components of the calculation of the combined $P({\Rate})$ considering the three observed DNS systems in the Galactic disk. The merger rate of a given population {\em i} can be defined by 

\begin{equation}
R_{i} \equiv \left(\frac{N_{\rm PSR}}{\tau_{\rm life}}f_{\rm b} \right)_{i} ~,
\end{equation}

where $N_{\rm PSR, i}$ represents the number of pulsars in our Galaxy with pulse and orbital characteristics {\it similar\/} to an observed sample {\em i} (e.g. PSR J0739$-$3039) and $f_{\rm b, i}$ is a correction factor to take into account pulsar beaming (typically $\sim 6$)\footnote{This is based on polarimetry measurements of B1913+16 and B1534+12.}. $\tau_{\rm life, i}$ is the lifetime of system {\em i} based on its observed properties. In order to calculate $N_{\rm PSR, i}$, we distribute a large population of pulsars {\em all}~similar to the system {\em i}, in a model galaxy assuming spatial and luminosity distributions. Since pulsar luminosities are drawn from a distribution, the observed flux and estimated distance for the DNS system are not relevant in our calculation. Moreover, the selection effects for faint pulsars are implicitly taken into account. Once we generate a pulsar population with a size $N_{\rm PSR}$, we can then calculate the number of pulsars detected ($N_{\rm det}$) by large-scale pulsar surveys. We repeat the survey simulations with a detailed modeling of selection effects for observed DNS systems. For a fixed $N_{\rm PSR}$, we find that $N_{\rm det}$ follows a Poisson distribution, P($N_{\rm det}; \bar N_{\rm det}$), where $\bar N_{\rm det}$ is a mean value of $N_{\rm det}$ for a given population size ($N_{\rm PSR}$). We require  $N_{\rm det}=1$, i.e., we consider only one observed system, and calculate the best-fit value of $\bar N_{\rm det}$. With a wide range of $N_{\rm PSR}$, we find $\bar N_{\rm det} = cN_{\rm PSR}$ as expected where c is a constant. Applying Bayes' theorem to these results, we calculate a P($N_{\rm PSR}$), a PDF for the population size of a given system {\em i} knowing that there is one observation. The calculation $P(R)$ from $P(N_{\rm PSR})$ is straightforwad, and the full derivation and formula can be found in Appendix~A of 
Kim et al.\ (2004)\nocite{nswd}.

The lifetime of a DNS system is defined by $\tau_{\rm life} \equiv
\tau_{\rm sd} + \tau_{\rm mrg}$, where $\tau_{\rm sd}$ is the spin-down
age of a recycled pulsar \citep{acw99} and $\tau_{\rm mrg}$ is the remaining lifetime until the two neutron stars merge \citep{pm63}.
Based on the most recent observations, we estimate the lifetime of J0737$-$3039 to be $\sim$230\,Myr \citep{double}. This is the shortest among the three observed systems. The beaming correction factor $f_{\rm b}$ is defined as the inverse of the fractional solid angle subtended by the pulsar beam. Its
calculation requires detailed geometrical information on the
beam. Following Kalogera et al.\ (2001) and updating the values with recent observations, 
we calcualte $f_{\rm b}$$\sim$5.7 for PSR B1913+16 \citep{wt02} and $\sim$6.0 for PSR B1534+12 \citep{sta04} Without good knowledge of the geometry of J0737$-$3039, we assume $f_{\rm b, J0737}$ to be the average value of the other two systems ($\simeq 5.9$).

\begin{figure}
\includegraphics[width=7.3cm]{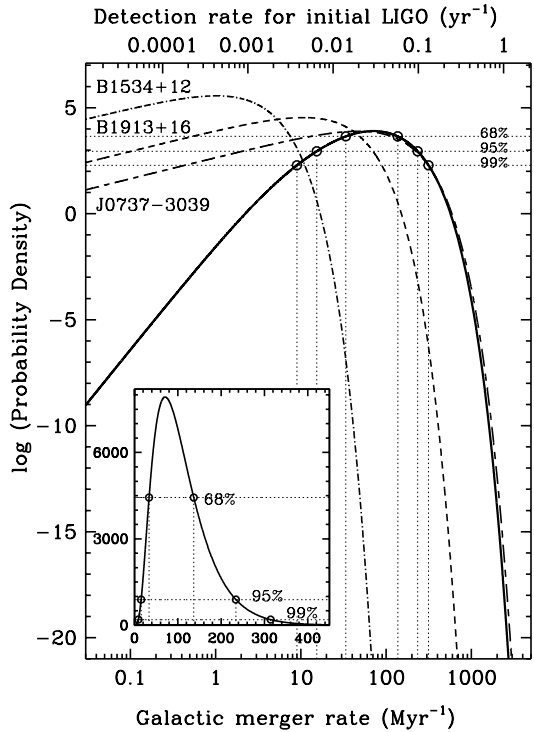}
\vspace{.6cm}
\caption{$P({\Rate})$ is shown on a log
scale. The thick solid line is the Galactic rate estimate
overlapped with results for individual observed systems (dashed
lines). Dotted lines indicate confidence intervals for the rate
estimates. The same results are shown on a linear scale in the small
inset. All results are from our reference model.}
\end{figure}

In Fig.\ 1, we show $P({\Rate})$ for a reference model (Model~6 in KKL\nocite{kkl}). We obtain the most likely value of ${\Rate} \sim 71\,{\rm Myr}^{-1}$,
larger by a factor of $\simeq 5.5$ than the rate estimated before the
discovery of J0737$-$3039. The increase factor is found similar for all
pulsar population models we examined. The increased merger rates imply a boost in the inferred detection rate of GWs from DNS inspirals for ground-based GW interferometers such as LIGO. In order to calculate the detection rate ($D$), we assume a homogeneous distribution of galaxies in nearby Universe and a spherical symmetry in detector sensitivity. Then, we can write ${\rm D} \equiv R_{\rm gal} \times N_{\rm gal}$, where $N_{\rm gal}$ is the number of galaxies in the detection volumne ($V_{\rm det}$). We calculate the number density of galaxies derived by the observed blue luminosity density, ($n_{\rm gal} = 1.25\times 10^{-2}$Myr$^{\rm -1}$ (see Phinney (1991) and Kalogera et al.\ (2001) \nocite{ph91,knst} for more details). The detection volume of LIGO can then be defined as a sphere for a given detection distance (20 Mpc and 350 Mpc for the inital and advanced LIGO, respectively), and the number of galaxies within $V_{\rm det}$ is simply $n_{\rm gal} \times V_{\rm det}$.  For our reference model, we find that the most probable event rates are about 1 per 30\,yrs and 1 per 2 days, for initial and advanced LIGO, respectively. At the 95\% confidence interval, the most optimistic predictions for the reference model are 1 event per 9\,yrs and 1.6 events per day for initial and advanced LIGO, respectively. More details can be found in Kalogera et al.\ (2004)\nocite{k04}.

As shown in the Fig.\ 1, the Galactic DNS merger rate is dominated by PSR~J0737$-$3039. Recently, Lorimer et al.\ (2005a)\nocite{l05} suggested that the current age of J0737$-$3039 is $\sim 30-70$ Myr. This implies a even shorter lifetime ($\tau_{\rm life} \sim115-155$ Myr) knowing that the estimated merger timescale of this system is $\sim$85 Myr. Based on their results, we find the most likely value of $\Rate$ for the reference model is $\simeq 90-110\,$Myr$^{-1}$.

The beaming correction for J0737$-$3039 is not yet constrained and we assume MSPs discovered in DNS systems are not very different. As a conservative lower limit, without any beaming corrections for all observed systems for a reference model, we obtain $\sim 12^{+11}_{-6}$ Myr$^{-1}$ with a 95\% confidence interval. The corresponding detection rates for initial and advanced LIGO are $5^{+12}_{-4} \times 10{^-3}$ yr$^{-1}$ and $27^{+62}_{-21}$ yr$^{-1}$, respectively. Only when the axis geometry of J0737-3039 becomes available, we will be able to constrain the uncertainties of the beaming fraction, and in turn, the rate estimates. 

\section{Global probability distribution of the rate estimates}

In KKL\nocite{kkl}, we showed that empirical DNS merger rates are strongly dependent on the assumed luminosity distribution function for pulsars, but not on the pulsar spatial distribution. Therefore, we can consider only the rate dependence on the pulsar luminosity function for simplicity. Here, we describe how we can incorporate the systematic uncertainties from these models and calculate, $P_{\rm g} ({\Rate})$, a {\em global\/} PDF of rate estimates. Note that the results available on prior functions for pulsar luminosity distribution are currently out of date. Specific quantitative results could change when constraints on the luminosity function are derived from the current pulsar sample.

We assume a power-law luminosity distribution for a radio pulsar luminosity function $f(L)$. This function is defined by two parameters: the cut-off luminosity $L_{\rm min}$ and power-index $p$. We assume prior distributions for these two parameters and calculate $P_{\rm g}({\Rate})$. We fit the marginalised likelihood of $L_{\rm min}$ and $p$ presented by \citep{cc97} and obtain the following analytic formulae for prior functions, i.e. $f(L_{\rm min})$ and $g(p)$: $f(L_{\rm min}) = \alpha_{\rm 0} + \alpha_{\rm 1} L_{\rm min} + \alpha_{\rm 2} L_{\rm min}^{2}$ and $g(p) = 10^{\beta_{\rm 0} + \beta_{\rm 1} p + \beta_{\rm 2} p^{2}}$, where $\alpha_{\rm i}$ and $\beta_{\rm i} ~({\rm i}=0,1,2)$ are coefficients we obtain from the least-square fits and the functions are defined over the intervals $L_{\rm min}=[0.0,\,1.7]$ mJy kpc$^{2}$ and $p=[1.4,\, 2.6]$. We note that, although Cordes and Chernoff (1997) obtained $f(L_{\rm min})$ over $L_{\rm min}\simeq [0.3,\,2]$ mJy kpc$^{2}$ centered at 1.1 mJy kpc$^{2}$, we consider $f(L_{\rm min})$ with a peak at $\sim 0.8\,$mJy kpc$^{2}$ considering the discoveries of faint pulsars with L$_{\rm 1400}$ below 1 mJy kpc$^{2}$ \citep{c03}. Next we calculate $P_{\rm g}({\Rate})$ as follows: $P_{\rm g}({\Rate})$\\
$= \int_{p}dp \int_{L_{\rm min}}dL_{\rm min} P(R)f(L_{\rm min})g(p)$.

\begin{figure}
\includegraphics[width=7.3cm]{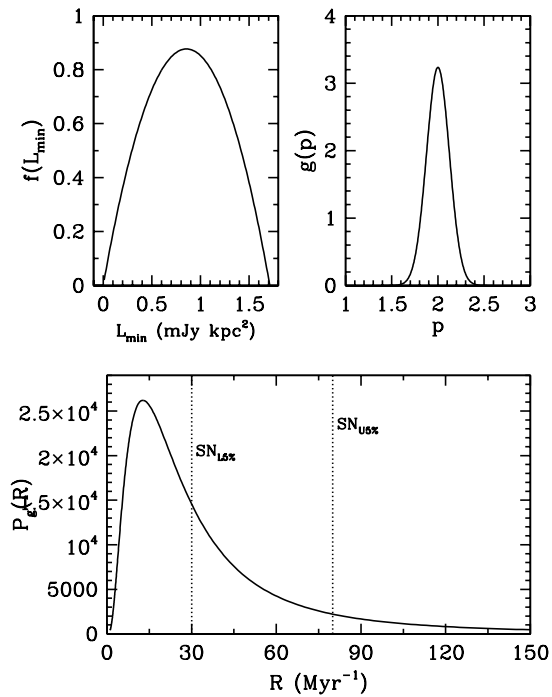}
\vspace{.6cm}
\caption{The global $P_{\rm g}({\Rate})$ on a linear scale (lower panel) and the assumed intrinsic distributions for $L_{\rm min}$ and $p$ (upper panels). Dotted lines represent the lower ($SN_{\rm L}$) and upper ($SN_{\rm U}$) bounds on the observed SN Ib/c rate scaled by 5\% of the observed SN Ib/c rates, $600-1600$ Myr$^{-1}$ (see text).}
\end{figure}

In Fig.\ 2, we show the distributions of $L_{\rm min}$ and $p$ adopted (top panels) and the resulting global distribution of Galactic DNS merger rate estimates (bottom panel). We find the peak value of $P_{\rm g}({\Rate})$ at {\em only\/} around 13\,Myr$^{-1}$. We note that this is a factor $\sim 5.5$ smaller than the result from the reference model (${\Rate}\sim$71\,Myr$^{-1}$). At the 95\% confidence interval, we find that the {\it global} Galactic DNS merger rates lie in the range $\sim$ 1--145\,Myr$^{-1}$. These imply LIGO event rates in the range $\sim (0.4-60)\times 10^{-3}\,$yr$^{-1}$ (initial) and $\sim 2-330\,$yr$^{-1}$ (advanced). Since 1997, the number of known millisecond pulsars has more than doubled, and therefore, constraints on $L_{\rm min}$ and $p$ and their PDFs based on the most up-to-date pulsar sample are urgently needed.

\section{Upper Limit of DNS Merger Rate Estimates. Constraints from Type Ib/c Supernovae Rates}
According to the standard binary evolution scenario, 
the progenitor of the second neutron star is expected to form during a Type Ib/c supernova. 
Therefore, the empirical estimates for the Type Ib/c SN rate in our Galaxy can be used to set upper limits on the DNS merger rate estimates. Based on Cappellaro et al.\ (1999)\nocite{cet99}, we adopt ${\Rate}_{\rm SN\,Ib/c} \simeq 1100\pm500\,$Myr$^{-1}$ considering Sbc--Sd galaxies, a Hubble constant $H_0=71\,$km/s/Mpc and the blue luminosity of our Galaxy $L_{\rm B,gal}=9\times10^{9}\,L_{\rm B,sun}$.

We calculate the fraction of SN Ib/c actually involved in the formation of DNS with a binary evolution code {\tt StarTrack} \citep{bkb02,bel05} and estimate the rate ratio: $\gamma\equiv({\Rate}_{\rm DNS}$/${\Rate}_{\rm SN\,Ib/c})\times100\le5\%$. 
Motivated by this result, we adopt the empirical ${\Rate}_{\rm SN\,Ib/c}$ assuming $\gamma\sim5\%$ and compare the value with the global PDF (Fig. 2).

\begin{table*}
\caption{Observed properties of tight DNS systems}
\label{defparagcl} 
\begin{center}
\begin{tabular}{l r r l l l}
\hline

PSR name & $P_{\rm s}^{a}$ (ms) & $P_{\rm b}^{b}$ (hr) &  e$^{c}$  &   $\tau_{\rm life}^{d}$ (Gyr)& $N_{\rm PSR}^{e}$  \\ 
\cline{1-6}
\hline
 B1913+16    &  59.03 &  7.75  &  0.617  &      0.37     & 680\\
 B1534+12    &  37.90 & 10.10  &  0.274  &      2.93     & 480\\
 J0737-3039A &  22.70 &  2.45   &  0.088  &      0.23     & 1680\\
 J1756-2251  &  28.46 &  7.67   &  0.181  &      2.03     & 400$-$600$^{f}$\\
 J1906+0746$^{g}$  & 144.14 &  3.98   &  0.085  &    0.082$^{h}$  & 300\\
\hline

\end{tabular}
\end{center}

\vspace*{.6cm}
\noindent
$^{a}$ pulsar spin period\\ 
$^{b}$ orbital period\\ 
$^{c}$ eccentricity\\ 
$^{d}$ lifetime of a system\\ 
$^{e}$ population size for a given DNS system\\ 
$^{f}$ approximate $N_{\rm pop}$. Results depend on the survey modeling (see text).\\ 
$^{g}$ The nature of the companion is not yet clear.
$^{g}$ $\tau_{\rm life}=\tau_{c}+\tau_{d}$ where $\tau_{\d}$ is a death-time of the pulsar.\\ 
\end{table*}

We note that our most optimistic DNS merger rate is ${\Rate}=189^{+691}_{-166}$ Myr$^{-1}$ at a 95\% confidence interval (Model~15 in KKL\nocite{kkl}). We obtain $\gamma \sim 80\%$ with respect to the center value of the empirical SN Type Ib/c rate ($1100$ Myr$^{-1}$) and the upper limit of ${\Rate}$ at the 95\% confidence interval ($189+691=880$ Myr$^{-1}$). This corresponds to $\gamma \sim 13\%$ for a SN Type II rate, which is factor 6.1 larger than that of SN Type Ib/c. In both cases, the most optimistic model is lower than the current empirical supernova rate estimates, but not really consistent with the results of population synthesis calculations. If we consider a upper limit of ${\Rate}$ at the 95\% confidence interval from the global PDF, we obtain $\gamma\sim$ 13\% and 2\% for the center value of SN Type Ib/c (1100 Myr$^{-1}$) and II, respectively.

\section{Implications of New Discoveries to the Galactic DNS merger rate estimates}

Recently, Faulkner et al.\ (2005) discovered PSR J1756$-$2251, the 4th merging DNS in the Galactic disk from the Parkes Multibeam Pulsar Survey (PMPS). The standard Fourier method failed to find this pulsar and they reanalysed the PMPS data with an acceleration search (or a `stack search' as described in their paper). In order to calculate the merger rate including PSR J1756$-$2251, a detailed simulation is necessary to calculate the effect of the acceleration search with the PMPS. However, the approximate contribution of PSR J1756$-$2251 to the Galactic merger rate can be easily obtained. We find the total rate increases by only $\sim4\%$ due to the new discovery. This is expected because PSR J1756$-$2251 can be identified as a member of the B1913+16-like population, which has already been taken into account in the calculation. Only future detections of pulsars from a significantly different population (compared to the known systems), or from the most relativistic systems, will result in a non-trivial contribution to the rate estimates. 

Finally, we note the implications of J1906+0746 on the pulsar binary merger rates. This system has drawn attention due to the extremely young age of the pulsar (characteristic age of $\sim$ 112 kyr; Lorimer et al.\ 2005b\nocite{j1906}). If the companion is another neutron star, J1906+0746 would be the first discovery of a non-recycled component in a DNS system. Currently, the nature of the companion is totally unknown, and it can be either a light neutron star, or heavy (O-Mg-Ne) white dwarf. Assuming J1906+0746 is a DNS system, we calculate its contribution to the Galactic merger rate. Because of its short lifetime ($\sim$82Myr), J1906+0746 can increase the Galactic DNS merger rate by about a factor 2. This implies that the current estimated DNS merger rate including J0737$-$3039 can still be doubled! If J1906+0746 is an eccentric NS--WD system, such as J1141-6545 \citep{vk99} or B2303+46 \citep{std85}, it will be as important as J1141-6545, which is currently dominate the birthrate of eccentric NS-WD binaries. In Table 1, we summarize some important properties of observed merging DNS systems, which used in the calculation of $P(\Rate)$ in this paper.

\section{Conclusion}

We summarize our results on the Galactic merger rate of double neutron stars (DNS) in view of the recent discovery of PSR J0737-3039. We update the results based on the most recent observations. The {\em global} PDF of rate estimates incorporates the already known systematics due to the radio pulsar luminosity function. The most likely value obtained from the global distribution is at only $\sim$ 13 Myr$^{-1}$, and a reanalysis of the {\em current} pulsar sample and radio luminosities is needed for a reliable assessment of the best fitting distribution. The DNS merger rate can be constrained by empirical Type Ib/c supernova rates, and we compare the supernova rate with the global PDF. Recent discoveries of relativistic systems, such as J1756$-$2251 and J1906+0746 are important for the reduction of the uncertainties associated with the empirical rate estimates. As for J1756$-$2251, we expect their contribution to the Galactic DNS merger rate is negligible, and this pulsar can be seen another sample of B1913+16-like population. It is not clear whether the companion of J1906+0746 is a neutron star or a white dwarf. This system may increase the estimated merger rate of the relevant population by a factor of 2 whether it is a DNS or another eccentric NS-WD binary.

\section{Acknowledgement}

This research is partially supported by NSF grant 0353111, and a Packard Foundation Fellowship in Science and Engineering to VK. DRL is a University Research Fellow supported by the Royal Society.

\end{document}